\documentclass[conference]{IEEEtran}

\IEEEoverridecommandlockouts
\usepackage{cite}
\usepackage{amsmath,amssymb,amsfonts}
\usepackage{algorithmic}
\usepackage{parcolumns}
\usepackage{graphicx}
\usepackage{textcomp}
\usepackage{xcolor}
\usepackage{gensymb}
\usepackage{tabularx}
\newcommand\numberthis{\addtocounter{equation}{1}\tag{\theequation}}
\def\BibTeX{{\rm B\kern-.05em{\sc i\kern-.025em b}\kern-.08em
    T\kern-.1667em\lower.7ex\hbox{E}\kern-.125emX}}

\begin{document}

\bstctlcite{IEEEexample:BSTcontrol}

\newcommand{\heading}[1]{\colchunk[1]{\hspace*{-\parindent}\textit{#1}}}
\newcommand{\desc}[1]{\colchunk[2]{#1}\colplacechunks}

\title{Favorable Propagation with User Cluster Sharing
%\thanks{The work of M. Matthaiou was supported by EPSRC, UK, under grant EP/P000673/1.} %%% Funding source goes here
}
\author{\IEEEauthorblockN{%
Chelsea Miller\IEEEauthorrefmark{1},
Peter J. Smith\IEEEauthorrefmark{2},
Pawel A. Dmochowski\IEEEauthorrefmark{1},
Harsh Tataria\IEEEauthorrefmark{3}, and
Andreas F. Molisch\IEEEauthorrefmark{4}}

   \IEEEauthorblockA{\IEEEauthorrefmark{1} School of Engineering and 
Computer Science, Victoria
  University of Wellington, Wellington, New Zealand}
   \IEEEauthorblockA{\IEEEauthorrefmark{2}School of Mathematics and 
Statistics, Victoria
  University of Wellington, Wellington, New Zealand}
	\IEEEauthorblockA{\IEEEauthorrefmark{3}Department of Electrical and Information Technology,
	Lund University, Lund, Sweden}
   \IEEEauthorblockA{\IEEEauthorrefmark{4}Department of Electrical Engineering,
  University of Southern California, Los Angeles, CA, U.S.A.}
   \IEEEauthorblockA{e-mail:~\{chelsea.miller, peter.smith, pawel.dmochowski\}@ecs.vuw.ac.nz,~harsh.tataria@eit.lth.se, and~molisch@usc.edu}
}

% \author{\IEEEauthorblockN{1\textsuperscript{st} Chelsea Miller}
% \IEEEauthorblockA{\textit{School of Engineering and Computer Science} \\
% \textit{Victoria University of Wellington}\\
% Wellington, New Zealand \\
% chelsea.miller@ecs.vuw.ac.nz}
% \and
% \IEEEauthorblockN{2\textsuperscript{nd} Pawel Dmochowski}
% \IEEEauthorblockA{\textit{School of Engineering and Computer Science} \\
% \textit{Victoria University of Wellington}\\
% Wellington, New Zealand \\
% \and
% \IEEEauthorblockN{3\textsuperscript{rd} Peter Smith}
% \IEEEauthorblockA{\textit{School of Mathematics and Statistics} \\
% \textit{Victoria University of Wellington}\\
% Wellington, New Zealand \\
% \and
% \IEEEauthorblockN{4\textsuperscript{th} Harsh Tataria}
% \and
% \and

\maketitle

\begin{abstract}
We examine the favorable propagation (FP) behavior of a massive multi-user multiple-input-multiple-output (MU-MIMO) system equipped with a uniform linear array (ULA), horizontal uniform rectangular array (HURA) or uniform circular array (UCA) using a ray-based channel model with user cluster sharing. We demonstrate FP for these systems and provide analytical expressions for the mean-squared distance (MSD) of the FP metric from its large-system limit for each of the aforementioned topologies. We use these results to examine the detrimental effects of user cluster sharing on FP behavior, and demonstrate the superior performance of the ULA as compared to the UCA and the HURA with equal inter-element spacing. Although cluster sharing has a negative impact on FP for finite arrays, we additionally examine the asymptotic rate of convergence to FP as a function of array size and show that this rate is unchanged with or without user cluster sharing. 
\end{abstract}

%%\begin{IEEEkeywords}
%%%%%%  NOT IN A CONFERENCE PAPER
%%\end{IEEEkeywords}

\section{Introduction}\label{intro}
%NOTE: need more citations for channel hardening in rayleigh and het. corr. ricean
Favorable propagation (FP) is a key concept underpinning the potential of massive MIMO systems \cite{ngo_aspects_2014}. As the number of base station (BS) antennas grows without bound, FP describes the resulting mutual orthogonality of user channel vectors. It is of intrinsic interest as
a fundamental measure of interference \cite{ngo_aspects_2014} and is essential
for the effective performance of matched filtering (MF) \cite{ngo_aspects_2014}. While alternative, interference-cancelling techniques such as zero-forcing (ZF) and minimum mean squared error (MMSE) combining do not require FP to combat interference, their performance is still aided by FP. This is shown in \cite{JSTSP} where the performance of ZF and MMSE is analytically linked to a measure of FP.

FP is also an intrinsically valuable property for the practical realisation of massive MIMO systems. Algorithms are being designed based on the assumption that FP holds \cite{ghavami2018blind} which has been shown to simplify various aspects of radio resource management \cite{zheng2017channel}. We note that MF is still the most preferred multiuser processing choice
from an implementation viewpoint at the BS, since it only consumes a small fraction of
the total baseband field programmable gate array (FPGA) resource usage, needing no real-time optimization
routines. This is in stark contrast
to the more sophisticated matrix inversion-based processing techniques,
which require memory-consuming high-level synthesis to optimize and manage the
real-time dataflow, in order to meet the overall post-scheduler
processing latency requirements of 5G New Radio (5G-NR) systems. 
%This work is now mature, but suers from three fundamental problems.
%First, there are large gaps in our ability to provide accurate statistical models without implicitly using ray-based channels anyway.
%Secondly, any conclusions are by their nature linked to the statistical model parameters and not to the underlying physical and electromagnetic properties of the system. 

There exists a great amount of analysis on FP for multi-user MIMO (MU-MIMO) systems with \textit{classical statistical} channel models, ranging from simple i.i.d. Rayleigh fading \cite{ngo_aspects_2014}, to more complex heterogeneous correlated Ricean models \cite{matthaiou_does_2018}. While providing valuable insight, their conclusions are by their nature linked to the statistical model parameters and not to the underlying physical and electromagnetic properties of the system and environment. Furthermore, propagation measurements suggest that millimeter-wave (mmWave) channels are better described by \textit{spatial} or \textit{ray-based} models \cite{sangodoyin_cluster_2018,zhang2019favorable
%	 ,akdeniz, 
%	 ko2017millimeter
}. While explicitly incorporating the effects of different environmental properties and antenna topologies, the resulting complex structure of these models leads to often intractable analysis. Hence, a large portion of literature which examines FP with these physically-motivated models does so through simulation or ray-tracing \cite{
	%li2018performance,
	aslam_performance_2019,
%	kurras2018application,
%	cheng_channel_2018,
%	gao2015massive,
	aslam2018massive
}. For example, FP is examined via ray-tracing of a dense urban location for a uniform linear array (ULA), uniform circular array (UCA), and uniform rectangular array (URA) in \cite{aslam_performance_2019}, and for a ULA and UCA in \cite{aslam2018massive}. The latter additionally simulates the WINNER II channel model. The majority of works which provide a mathematical analysis rely on various assumptions around the angular distributions \cite{
wu_favorable_2017,
%masouros_space-constrained_2015,
%tan2018spectral,
zhang2019favorable,
%shehata2019theoretical,
yang2017massive,
gao_asymptotic_2015
}. The authors of \cite{wu_favorable_2017} examine FP conditions using a ray based model with uniform angles for a ULA, horizontal URA (HURA), and UCA, while \cite{yang2017massive} and \cite{gao_asymptotic_2015} do so for a ULA only. The analysis in\cite{zhang2019favorable} uses Gaussian and Laplacian distributed sub-rays, but still imposes the assumption of uniform cluster angles. In recent literature, there are only a handful of examples which provide generic closed-form analysis of MU-MIMO systems for ray-based models with arbitrary angular distributions \cite{chelsea_icc19, JSTSP, jasmine_icc19}.

To the best of the authors' knowledge, all of the analytical work to date applies exclusively to ray-based models in the absence of user cluster sharing, making the channels of different users statistically independent. However, models such as the COST 2100 model \cite{Oestges2012RadioCM} can generate a significant probability that a scattering cluster will be visible to multiple users. This prevalence of common scatterers is supported by measurement \cite{poutanen2010significance}. Although the significant impact of common scatterers on inter-user channel correlation (and thus FP) has been 
%\footnote{`...underestimating the significance of common scatterers in simulations would result in overestimating the system performance`\cite{Oestges2012RadioCM}} 
shown via simulation \cite{poutanen2012multi}, all analytical progress neglects this and is instead facilitated by the assumption that the user channels are independent. In anticipation of the potential effects of common clusters, we examine the FP condition for a MU-MIMO system with ray-based channels featuring user cluster sharing and a range of antenna topologies. More specifically, our contributions are as follows.
%CONTRIBUTIONS:
\begin{itemize}
	\item While the existence of FP has been proven for a uniform linear array (ULA) and a horizontal uniform rectangular array (HURA) \textit{without} cluster sharing in \cite{2019_TWC_SLi}, we prove that FP holds with cluster sharing for ULA and HURA configurations. We also conjecture that FP holds for a UCA with and without user cluster sharing and provide a mathematical basis for the conjecture which may have applications to more general topologies (see Sec.~\ref{s:FP}).
%	\item While closed-form solutions for the MSE of FP/CH metrics from their large system limits have been solved in \cite{JSTSP} for a ULA and HURA without cluster sharing, we derive the same for a ULA, HURA, and UCA with cluster sharing in Sec.~\ref{s:derivations}
	\item In Sec.~\ref{analysis}, we derive analytical expressions for the mean squared distance (MSD) of the FP metric from its large system limit for a finite-antenna system equipped with a ULA, HURA, and UCA with cluster sharing. We use this as a measure of the ``distance from FP" for a finite-antenna system.
%	\item Sec.~\ref{s:derivations} also contains corresponding closed-form solutions for the MSD of the CH metric from its large system limit.
	\item In Sec.~\ref{s:kappa_decay} we examine the ergodic rate of convergence to FP for a ULA with and without cluster sharing.
	\item We observe that user cluster sharing has a detrimental impact on FP behavior, while topologies with larger azimuth footprints promote FP behavior. We discuss how the superior spatial resolution of such topologies provides resilience to the correlated channel conditions caused by cluster sharing, reducing the MSD from FP.
\end{itemize}
%
%
%
%%%%%%%%%%%% SYSTEM MODEL %%%%%%%%%%%%%%%
\section{System Model}\label{systemmodel}
We consider uplink (UL) transmission in a single-cell MU-MIMO system where a base station (BS) is equipped with $M$ antennas and serves several single-antenna users (UEs).
%As analysis for generic ray-based models is in its infancy, we focus on a simplified initial case assuming: perfect CSI at the BS, and omnidirectional per-element antenna patterns.
%The link gain between the BS and the $l^\textrm{th}$ UE is determined through the classic pathloss and shadowing formula $\beta^{(l)} = AX_l\left(d_l/d_0\right)^{\Gamma}$, where $d_l$ is the distance between the BS and UE $l$, $\Gamma$ is the pathloss exponent, and $10\log_{10}(X_l/10)\sim\mathcal{N}(0,\sigma_\textrm{sf}^2)$ accounts for lognormal shadow fading. Also, $A$ is a unitless attenuation constant describing the power loss at reference distance $d_0$. This link gain is divided into $\beta_c^{(l)}$ for each cluster $c$ seen by UE $l$ in an exponentially decreasing manner. Each subray $s$ then has ray power $\beta_{c,s}^{(l)} = \beta_{c}^{(l)}/S$.
%\subsection{Channel Model}\label{ss:channel_model}
We adopt a clustered ray-based channel model. Within each \textit{drop}, the angular parameters for the UEs are defined by the following process.
%the UEs are randomly distributed within the cell area, outside of the BS exclusion zone of radius $r_0$, using a Poisson point process. 
$C_T$ scattering clusters are each assigned a random central azimuth angle of arrival (AAoA), $\phi_{c}$, and central elevation angle of arrival (EAoA), $\theta_{c}$, where $c = 1, 2, \dots C_T$. Each user $l$ is randomly allocated (with equal probability) a set of $C$ visible clusters, $\mathcal{C}^{(l)}$, where $C\leq C_T$ and $\mathcal{C}^{(l)}\subset\{1\dots C_T\}$. Each cluster $c$ scatters a user's signal into $S$ sub-rays with random, instantaneous angular offsets $\Delta_{c,s}^{(l)}$ and $\delta_{c,s}^{(l)}$ in azimuth and elevation, respectively (exact distributions are discussed later in the text). Hence the AAoA of ray $s$ from user $l$ through cluster $c$ is $\phi_{c,s}^{(l)} = \phi_{c} + \Delta_{c,s}^{(l)}$ and similarly the EAoA is $\theta_{c,s}^{(l)} = \theta_{c} + \delta_{c,s}^{(l)}$. While one or more central angles may be shared by multiple UEs for whom a common cluster is visible, the subray offsets are assumed to be i.i.d. as different UE locations will lead to different reflection points along a common scatterer. The resulting $M\times1$ channel for user $l$ is given by
\begin{align}\label{eq:channel_model}
\mathbf{h}_l = \sum_{c\in \mathcal{C}^{(l)}}\sum_{s = 1}^{S}\gamma_{c,s}^{(l)}\mathbf{a}\left(\phi_{c,s}^{(l)},\theta_{c,s}^{(l)}\right),
\end{align}
where $\gamma_{c,s}^{(l)} = \sqrt{\beta_{c,s}^{(l)}}e^{j\Theta_{c,s}^{(l)}}$ is the ray coefficient, $\Theta_{c,s}^{(l)}\sim\mathcal{U}(0,2\pi)$ are i.i.d. phase shifts modelling fast fading, and the ray powers, $\beta_{c,s}^{(l)}$, satisfy $\sum_{c\in \mathcal{C}^{(l)}}\sum_{s=1}^S \beta_{c,s}^{(l)} = \beta^{(l)}$. We define $\beta^{(l)}$ as the overall link gain between user $l$ and the BS, which is divided amongst the visible clusters such that the cluster powers, $\beta_c^{(l)}$, add up to $\beta^{(l)}$. Each subray $s$ has ray power $\beta_{c,s}^{(l)} = \beta_{c}^{(l)}/S$. Note that the large-scale fading simply scales the user channel vectors. It has no impact on the existence of FP, and as shown in Sec.~\ref{ss:kappa_FP}, it scales the MSD from FP. We  include $\beta^{(l)}$ in the analysis for completeness but set $\beta^{(l)}=1\hspace{0.1cm}\forall\hspace{0.1cm}l$ for all numerical results. This normalises the channel vectors.

The steering vector $\mathbf{a}(\phi,\theta)$ for an arbitrary ray with angles $\phi$ and $\theta$ is governed by the antenna topology at the receiver. For a ULA located along the $x$-axis with an antenna spacing of $d_x$ wavelengths, the $m^\textrm{th}$ entry of $\mathbf{a}(\phi,\theta)$ is given by \cite{wu_favorable_2017}
\begin{align}\label{eq:ULA_steering_vector}
\left(\mathbf{a}(\phi,\theta)\right)_m = e^{j2\pi(m-1) d_x \sin{\phi}}.
\end{align}
For a HURA situated in the azimuth plane with antenna spacings $d_x$ and $d_y$,
\begin{equation}\label{eq:HURA_steering_vector}
\mathbf{a}(\phi,\theta) = \mathbf{a}_x(\phi,\theta) \otimes \mathbf{a}_y(\phi,\theta).
\end{equation}
Let $M_x$ and $M_y$ be the number of antennas along the $x$- and $y$-axis respectively, then the entries of the $M_x \times 1$ vector $\mathbf{a}_x(\cdot)$ are defined as \cite{wu_favorable_2017}
\begin{align}
\left(\mathbf{a}_x(\phi,\theta)\right)_m = e^{j2\pi d_x (m-1) \sin{\theta} \cos{\phi}}
\end{align}
and those of the $M_y \times 1$ vector $\mathbf{a}_y(\cdot)$ are
\begin{align}
\left(\mathbf{a}_y(\phi,\theta)\right)_m = e^{j2\pi d_y (m-1) \sin{\theta}\sin{\phi}}.
\end{align}
Finally, the entries of the steering vector for a UCA in the azimuth plane with antenna spacing $d_r$ are given as \cite{wu_favorable_2017}
\begin{align}\label{eq:UCA_steering_vector}
\left(\mathbf{a}(\phi,\theta)\right)_m = e^{j \frac{\pi d_r}{\sin(\pi/M)} \sin{\theta}\cos{\left(\phi - \psi_{m}\right)}},
\end{align}
where $\psi_m = 2\pi m/M$.
%\subsection{Cluster Sharing}\label{s:cluster_sharing}
We adopt the simple cluster sharing mechanism previously described to emulate that in the COST 2100 model - the most complete sharing mechanism which has been proposed thus far. In COST 2100, each cluster is allocated one or more visibility regions (VR); a cluster is said to be visible to a user if that user falls within the cluster's VR(s). A cluster can thus be visible to, and hence shared by, multiple UEs. Our model provides a simple mechanism to introduce this feature. Analysis of cluster sharing only depends on the probability that a randomly selected ray from one user originates from the same cluster as a randomly selected ray from another user. This sharing probability is denoted by $p_{\textrm{sh}}$ and can be simply adjusted by changing $C_T$ as the sharing mechanism gives $p_{\textrm{sh}}=1/C_T$. Although this mechanism limits the possible values of $p_\textrm{sh}$ in simulations, it provides a much simpler simulation method for the purpose of verifying analytical results which can then be used for any value of $p_\textrm{sh}$ between 0 and 1 in the analysis. 

Note that this model contains several approximations. For example, the visibility of clusters might be a function of the distance between the UE and the cluster. Additionally, a per-cluster shadowing might occur or, for very small distances between two UEs, the angles of the subrays within the cluster might become correlated. While incorporating these effects would further refine the channel model, it would make analytical treatment very difficult. The model we consider is thus a compromise, but importantly one that is more accurate than all the models previously used.

\section{Existence of FP with Cluster Sharing}\label{s:FP}
In this section we examine the existence of FP for ray-based channels with inter-user cluster sharing. Sec.~\ref{ss:FP} presents generic conditions for FP to hold with these models, while Sec.~\ref{s:FP_limits} examines these conditions for a ULA, HURA, and UCA. Recall that FP has been proven in \cite{2019_TWC_SLi} for a ULA and HURA in the absence of cluster sharing. 
Here, we extend these results by considering shared clusters. We also conjecture the existence of FP for a UCA with and without cluster sharing and provide a posible methodology for a mathematical proof.
\subsection{Requirements for FP}\label{ss:FP}
%{\color{blue}
%\begin{itemize}
%	\item Define Favorable Propagation (FP) in words
%	\item Define FP as an inner product of different users' channel vectors
%	\item Show FP convergence for ULA, HURA, UCA
%	\item Look at rate of FP convergence in large system limit
%\end{itemize}}
FP requires $\lim_{M\to\infty}\textbf{h}_{l}^\textrm{H}\textbf{h}_{l'}/M \to 0\hspace{0.3cm}\forall\hspace{0.3cm}l\neq l'$. Substituting the channel model from \eqref{eq:channel_model}, $\textbf{h}_{l}^\textrm{H}\textbf{h}_{l'}$ becomes
\begin{align}
\textbf{h}_{l}^\textrm{H}\textbf{h}_{l'}&=\sum_{\substack{s,c\\\in \mathcal{C}^{(l)}}}\sum_{\substack{s',c'\\\in \mathcal{C}^{(l')}}} \gamma^{(l)*}_{c,s}\gamma^{(l')}_{c',s'}\textbf{a}^\textrm{H}(\phi^{(l)}_{c,s},\theta^{(l)}_{c,s})\textbf{a}(\phi^{(l')}_{c',s'},\theta^{(l')}_{c',s'})\notag\\
&=\sum_{{s,c\in \mathcal{C}^{(l)}}}\sum_{\substack{s',c'\in \mathcal{C}^{(l')}\\c'\neq c}}\gamma^{(l)*}_{c,s}\gamma^{(l')}_{c',s'}\textbf{a}^\textrm{H}(\phi^{(l)}_{c,s},\theta^{(l)}_{c,s})\textbf{a}(\phi^{(l')}_{c',s'},\theta^{(l')}_{c',s'})\notag\\
&+ \sum_{\substack{s,c\\\in \mathcal{C}^{(l)}}}\sum_{s'=1}^{S} \gamma^{(l)*}_{c,s}\gamma^{(l')}_{c,s'}\textbf{a}^\textrm{H}(\phi^{(l)}_{c,s},\theta^{(l)}_{c,s})\textbf{a}(\phi^{(l')}_{c,s'},\theta^{(l')}_{c,s'})\notag\\
&\triangleq T_1 + T_2,
\end{align}
where $\sum_{s,c\in \mathcal{C}^{(l)}}$ denotes $\sum_{c\in \mathcal{C}^{(l)}}\sum_{s=1}^{S}$. The existence of FP requires $\lim_{M\to\infty}T_1/M \to 0$ and $\lim_{M\to\infty}T_2/M \to 0$, where $T_2$ isolates cases of shared clusters between users $l$ and $l'$.
\subsection{FP with Different Antenna Configurations}\label{s:FP_limits}
The existence of FP for ULA, HURA, and UCA antenna configurations can be determined by examining the limits of $T_1$ and $T_2$ for the steering vectors defined in \eqref{eq:ULA_steering_vector}, \eqref{eq:HURA_steering_vector}, and \eqref{eq:UCA_steering_vector}.
\subsubsection{ULA and HURA}\label{sss:FP_ULA}
In \cite{2019_TWC_SLi}, it is proven that $\lim_{M\to\infty}\frac{1}{M}T_1 \xrightarrow{a.s.} 0$ for a ULA and HURA for a generic ray based model, without cluster sharing. The analysis in \cite{2019_TWC_SLi} hinges on the results
\begin{align}\label{ULAFP_breaks}
P(\sin\phi_{c,s}^{(l)} = \sin\phi_{c',s'}^{(l')}) = 0
\end{align}
for a ULA, and
\begin{align}
P(\sin\theta_{c,s}^{(l)}\cos\phi_{c,s}^{(l)} = \sin\theta_{c',s'}^{(l')}\cos\phi_{c',s'}^{(l')})=0,\label{HURAFP_breaks1}\\
P(\sin\theta_{c,s}^{(l)}\sin\phi_{c,s}^{(l)} = \sin\theta_{c',s'}^{(l')}\sin\phi_{c',s'}^{(l')})=0\label{HURAFP_breaks2}
\end{align}
for a HURA. In \cite{2019_TWC_SLi}, these results followed from the fact that the angles $\phi_{c,s}^{(l)}$, $\theta_{c,s}^{(l)}$, $\phi_{c',s'}^{(l')}$, $\theta_{c',s'}^{(l')}$ are continuous random variables chosen independently amongst clusters, sub-rays, and users. Exactly the same argument applies to  $T_1$ as here the rays are independent. With shared clusters, the situation is different as the rays in $T_2$ are dependent due to the presence of a shared cluster ($c=c'$). Nevertheless, the probabilities in \eqref{ULAFP_breaks}, \eqref{HURAFP_breaks1}, and \eqref{HURAFP_breaks2} are still zero as the i.i.d. angular offsets make the angles $\phi_{c,s}^{(l)}$, $\theta_{c,s}^{(l)}$, $\phi_{c',s'}^{(l')}$, $\theta_{c',s'}^{(l')}$ conditionally independent continuous random variables, where the conditioning is on the central angles. Hence, FP holds for a ULA and HURA configuration with and without cluster sharing.
%In addition, using the same argument involving \eqref{ULAFP_breaks}, the authors of {\color{blue}Jasmine journal} prove that the rate of convergence to FP for a ULA is $\mathcal{O}(1/M)$. Hence, we can conclude that the same is true for the clustered ray-based model with cluster sharing.
\subsubsection{UCA}
%{\color{blue}We need to tidy this up!}
Using the UCA steering vector in \eqref{eq:UCA_steering_vector}, for FP we have
\begin{align*}
T_1 &+ T_2= \sum_{\substack{\vspace{0.0075cm}\\s,c\\\in \mathcal{C}^{(l)}}}\sum_{\substack{s',c'\\ \in \mathcal{C}^{(l')}}}\gamma^{(l)*}_{c,s}\gamma^{(l')}_{c',s'}\\
&\times\sum_{m=0}^{M-1} e^{\frac{-j\pi d_r}{\sin(\pi/M)}[\sin\theta_{c,s}^{(l)}\cos(\phi_{c,s}^{(l)} - \psi_m)-\sin\theta_{c',s'}^{(l')}\cos(\phi_{c',s'}^{(l')} - \psi_m)]}\\
&=
\sum_{\substack{\vspace{0.0075cm}\\s,c\\\in \mathcal{C}^{(l)}}}\sum_{\substack{s',c'\\ \in \mathcal{C}^{(l')}}}\gamma^{(l)*}_{c,s}\gamma^{(l')}_{c',s'}\sum_{m=0}^{M-1} e^{\frac{-j\pi d_r}{\sin(\pi/M)} \sqrt{a^2 + b^2}\sin(\psi_m + \alpha)}\\
%&\approx
%\sum_{\substack{\vspace{0.0075cm}\\c,s\\\in \mathcal{C}^{(l)}}}\sum_{\substack{c',s'\\ \in \mathcal{C}^{(l')}}}\gamma^{(l)*}_{c,s}\gamma^{(l')}_{c',s'}\sum_{m=0}^{M-1} e^{-jMd_r\sqrt{a^2 + b^2}(\sin(\psi_m + \alpha))}\numberthis\label{eq:UCA_FP_1}
\end{align*}
where $a = \sin\theta_{c,s}^{(l)}\cos\phi_{c,s}^{(l)} - \sin\theta_{c',s'}^{(l')}\cos\phi_{c',s'}^{(l')}$, $b = \sin\theta_{c,s}^{(l)}\sin\phi_{c,s}^{(l)} - \sin\theta_{c',s'}^{(l')}\sin\phi_{c',s'}^{(l')}$, and $\alpha = \tan^{-1}(a/b)$. Hence, FP holds if
\begin{align}\label{eq:UCA_FP_condition}
\lim_{M\to\infty}\frac{1}{M}\sum_{m=0}^{M-1} e^{-jMd_r\sqrt{a^2 + b^2}\sin(\psi_m + \alpha)} \to 0
\end{align}
%For CH, we have 
%\begin{align*}
%T_4\hspace{-0.1cm}=\hspace{-0.1cm}\sum_{\substack{\vspace{0.0075cm}\\c,s\\\in \mathcal{C}^{(l)}}}\hspace{-0.2cm}\sum_{\substack{c',s'\\ \in \mathcal{C}^{(l)}\\\hspace{-0.7cm}\{c,s\}\neq\{c',s'\}}}\hspace{-0.2cm}\gamma^{(l)*}_{c,s}\gamma^{(l)}_{c',s'}\sum_{m=0}^{M-1} &e^{jMd_r\sin\theta_{c,s}^{(l)}\cos(\phi_{c,s}^{(l)} - \psi_m)}\numberthis\label{eq:UCA_CH_1}\\
%\times &e^{-jMd_r\sin\theta_{c',s'}^{(l)}\cos(\phi_{c',s'}^{(l)} - \psi_m)},
%\end{align*}
%again using the small angle approximation. Hence, for CH we require $T_4 \to 0$ or $T_4 \to \textrm{X}$. Note that \eqref{eq:UCA_CH_1} is simply \eqref{eq:UCA_FP_condition} with $l' = l$.
%
%Both CH and FP hold for a UCA if it can be shown that \eqref{eq:UCA_FP_condition} holds.
where we utilise the small angle approximation $\sin(\pi/M) \approx \pi/M$ for large $M$. At this stage, the convergence remains a conjecture, supported by numerical results in Sec.~\ref{results}. Intuitively, convergence occurs because the complex exponential in \eqref{eq:UCA_FP_condition} is averaged over $M$ values of the complex argument ranging from $-Md_r\sqrt{a^2+b^2}$ to $Md_r\sqrt{a^2+b^2}$. This large number of points wraps around the unit circle many times, becoming nearly uniform on $[0,2\pi]$ for large $M$ so that the average converges to zero. This is the idea behind several mathematical proofs, for example \cite{Koksma}. Unfortunately, established results do not precisely cover the UCA sum in \eqref{eq:UCA_FP_condition} for two reasons. First, the sum in \eqref{eq:UCA_FP_condition} is not a sum of fixed terms as $M \to \infty$ as the terms are actually functions of $M$. Secondly, most proofs rely on the exponential arguments forming a lacunary sequence \cite{Koksma}, meaning that the arguments can be ordered so there is a minimum ratio between adjacent terms that is greater than 1. This is not the case in \eqref{eq:UCA_FP_condition} where two arguments can be arbitrarily close to each other. These technical problems with a proof have been relaxed in various similar situations. Hence, the result remains a conjecture, albeit a well-motivated one. This conjecture may have extensive applications as a proof based on this argument may be the basis of a proof for any practical antenna topology.
%{\color{red}TODO: hopefully show that this holds for shared clusters as well!}
%using the properties of ray phases detailed in \eqref{eq:ray_phase_properties}
\section{Finite System Analysis}\label{analysis}
In this section, we examine the MSD of $\mathbf{h}_{l}^\textrm{H}\mathbf{h}_{l'}/M$ from the FP limit as a means of assessing the proximity to FP conditions for finite-antenna systems. Sec.~\ref{ss:kappa_FP} provides a generic analysis of this metric for ray-based models with user cluster sharing, while Sec.~\ref{s:derivations} provides analytical expressions for a ULA, HURA, and UCA. Note that a similar distance metric is used in \cite{wu_favorable_2017}, while alternative metrics can be found in, for example, \cite{ngo_aspects_2014}.
\subsection{Mean Squared Distance from FP}\label{ss:kappa_FP}
We examine $\kappa^\textrm{FP}$ which we define as the MSD of $\mathbf{h}_{l}^\textrm{H}\mathbf{h}_{l'}/M$ from the FP limit. Having proven the existence of FP in Sec.~\ref{s:FP_limits}, $\kappa^{\text{FP}}$ is given by
\begin{align}\label{eq:kappa_FP_2}
\kappa^\textrm{FP} &= \frac{1}{M^2}\mathbb{E}[|\textbf{h}_{l}^\textrm{H}\textbf{h}_{l'}|^2].
\end{align}
Substituting \eqref{eq:channel_model}, and noting that $\mathbb{E}[\gamma_{c,s}^{(l)*}\gamma_{c',s'}^{(l')}] = 0$ for $c\neq c'$ or $s\neq s'$ or $l\neq l'$, we obtain
\begin{align*}
&\kappa^\textrm{FP} 
%&= \frac{1}{M^2}\mathbb{E}\left[\left|\sum_{\substack{c,s\\\in \mathcal{C}^{(l)}}} \sum_{\substack{c',s'\\ \in \mathcal{C}^{(l)}}}\gamma^{(l)*}_{c,s}\gamma^{(l')}_{c',s'}\mathbf{a}^\textrm{H}(\phi_{c,s}^{(l)})\mathbf{a}(\phi_{c',s'}^{(l')})\right|^2\right]\\
%&= \frac{1}{M^2}\mathbb{E}[\sum_{\substack{c,s\\\in \mathcal{C}^{(l)}}} \sum_{c',s'\in \mathcal{C}^{(l')}}\sum_{\hat{c},\hat{s}\in \mathcal{C}^{(l)}} \sum_{\hat{c}',\hat{s}'\in \mathcal{C}^{(l')}}\gamma^{(l')*}_{\hat{c}',\hat{s}'}\gamma^{(l)}_{\hat{c},\hat{s}}\gamma^{(l)*}_{c,s}\gamma^{(l')}_{c',s'}\\
%&\times\mathbf{a}^\textrm{H}(\phi_{c,s}^{(l)})\mathbf{a}(\phi_{c',s'}^{(l')})\mathbf{a}^\textrm{H}(\phi_{\hat{c},\hat{s}}^{(l)})\mathbf{a}(\phi_{\hat{c}',\hat{s}'}^{(l')})]\\
= \frac{1}{M^2}\mathbb{E}\big[\sum_{\substack{s,c\\\in \mathcal{C}^{(l)}}} \sum_{\substack{s',c'\\\in \mathcal{C}^{(l')}}}|\gamma^{(l)}_{c,s}|^2|\gamma^{(l')}_{c',s'}|^2\\
&\times\mathbf{a}^\textrm{H}(\phi_{c,s}^{(l)},\theta_{c,s}^{(l)})\mathbf{a}(\phi_{c',s'}^{(l')},\theta_{c',s'}^{(l')})\mathbf{a}^\textrm{H}(\phi_{c',s'}^{(l')},\theta_{c',s'}^{(l')})\mathbf{a}(\phi_{c,s}^{(l)},\theta_{c,s}^{(l)})\big]\\
&= \beta^{(l)}\beta^{(l')}\left((1-p_\textrm{sh})K_c + p_\textrm{sh}K_s\right),\numberthis\label{eq:kappa_FP_3}
\end{align*}
where we define
\begin{align}\label{eq:Kc_definition}
K_c &= \frac{1}{M^2}\mathbb{E}[|\mathbf{a}^\textrm{H}(\phi_{c,s}^{(l)},\theta_{c,s}^{(l)})\mathbf{a}(\phi_{c',s'}^{(l')},\theta_{c',s'}^{(l')})|^2]
\end{align}
\begin{align}\label{eq:Ks_definition}
K_s &= \frac{1}{M^2}\mathbb{E}[|\mathbf{a}^\textrm{H}(\phi_{c,s}^{(l)},\theta_{c,s}^{(l)})\mathbf{a}(\phi_{c,s'}^{(l)},\theta_{c,s'}^{(l)})|^2].
\end{align}
Equation \eqref{eq:kappa_FP_3} follows from first taking the expectation inside the sum. Then, we note that the expectation over the ray angles only depends on whether the two rays share a cluster. The ray indexed by $c,s$ has the same cluster as the ray indexed by $c',s'$ with probability $p_{sh}$. Similarly the two rays have different clusters with probability $1-p_{sh}$.
In essence, the distance from FP for a finite antenna system hinges on the expected inner product of the steering vectors of two rays. $K_c$ and $K_s$ are exactly this, for the cases of unique cluster central angles and common cluster central angles, respectively. Here, we note that the large-scale fading values $\beta^{(l)}$ and $\beta^{(l')}$ simply scale $\kappa^\textrm{FP}$, and that $\kappa^\textrm{FP}$ is independent of the number of clusters, $C$, and subrays, $S$, which contribute to each UE's channel.
\subsection{Solutions for $K_c$ and $K_s$}\label{s:derivations}
\subsubsection{ULA}\label{ss:ULA_derivation}
%{\color{blue}\begin{itemize}
%	\item reference ICC for Kc and Ks
%	\item give Kc and Ks
%\end{itemize}}
Solutions for $K_c^\textrm{ULA}$ and $K_s^\textrm{ULA}$ are given in \cite{chelsea_icc19} as
\begin{align}\label{eq:Kc_ULA}
&K_c^\textrm{ULA} = \frac{1}{M^2}\sum_{m,m'}^{M-1}|\mathbb{E}[e^{j2\pi d_x(m-m')\sin\phi}]|^2\\\notag
&=\frac{1}{M^2}\sum_{m,m'}^{M-1}\left|\sum_{n = -\infty}^{\infty}\chi_c^\textrm{az}(n)\chi_s^\textrm{az}(n)J_n(2\pi d_x(m'-m))\right|^2,
\end{align}
and
\begin{align*}
&K_s^\textrm{ULA} = \frac{1}{M^2}\sum_{m,m'}^{M - 1}\mathbb{E}\big[e^{-j2\pi d_x(m-m')(\sin(\phi_{c,s}^{(l)}) - \sin\phi_{c,s'}^{(l')})}\big]\\\notag
&= \frac{1}{M^2}\sum_{m,m'}^{M-1}\bigg|\sum_{n = -\infty}^{\infty}\sum_{n' = -\infty}^{\infty}\chi_c^\textrm{az}(n-n')\chi_s^\textrm{az}(n)\chi_s^\textrm{az*}(n')\\
&\times J_n(2\pi d_x(m'-m))J_{n'}(2\pi d_x(m'-m))\bigg|^2,\numberthis\label{eq:Ks_ULA}
\end{align*}
%\begin{align}\label{eq:Kc_ULA_form}
%K_c^\textrm{ULA} &= \frac{1}{M^2}\mathbb{E}[|\sum_{m=0}^{M-1}e^{-j2\pi d_xm\sin\phi_{c,s}^{(l)}}e^{j2\pi d_xm\sin\phi_{c',s'}^{(l')}}|^2]\\\notag
%&= \sum_{m = 0}^{M - 1}\sum_{m = 0}^{M - 1}\mathbb{E}[|e^{j2\pi d_x(m-m')\sin\phi}|^2].
%\end{align}
%Similarly, from \eqref{eq:Ks_definition} we have
%\begin{align}\label{eq:Ks_ULA_form}
%K_s^\textrm{ULA} &= \frac{1}{M^2}\mathbb{E}[|\sum_{m=0}^{M-1}e^{-j2\pi d_xm\sin\phi_{c,s}^{(l)}}e^{j2\pi d_xm\sin\phi_{c,s'}^{(l')}}|^2]\\\notag
%&=\sum_{m = 0}^{M - 1}\sum_{m = 0}^{M - 1}\mathbb{E}[e^{-j2\pi d_x(m-m')(\sin\phi_{c,s}^{(l)} - \sin\phi_{c,s'}^{(l')})}].
%\end{align}
%From \cite{chelsea_icc19}, we have
%\begin{align}\label{eq:Kc_ULA}
%K_c^\textrm{ULA} &= \frac{1}{M^2}\sum_{m = 0}^{M-1}\sum_{m' = 0}^{M-1}\\\notag
%&\left|\sum_{n = -\infty}^{\infty}\chi_c^\textrm{az}(n)\chi_s^\textrm{az}(n)J_n(2\pi d_x(m'-m))\right|^2
%\end{align}
%\begin{align}\label{eq:Ks_ULA}
%K_s^\textrm{ULA} &= \frac{1}{M^2}\sum_{m = 0}^{M-1}\sum_{m' = 0}^{M-1}\\\notag
%&\bigg|\sum_{n = -\infty}^{\infty}\sum_{n' = -\infty}^{\infty}\chi_c^\textrm{az}(n-n')\chi_s^\textrm{az}(n)\chi_s^\textrm{az*}(n')\\\notag
%&J_n(2\pi d_x(m'-m))J_{n'}(2\pi d_x(m'-m))\bigg|^2,
%\end{align}
where we write $\sum_{i,j}^{I-1}$ in place of $\sum_{i = 0}^{I-1}\sum_{j = 0}^{I-1}$ and $J_{n}$ is the $n^\textrm{th}$ order Bessel function of the first kind. Here, $\chi_c^\textrm{az}(n) = \mathbb{E}_{\phi}[e^{jn\phi}]$ and $\chi_s^\textrm{az}(n) = \mathbb{E}_{\Delta}[e^{jn\Delta}]$ are the characteristic functions for the azimuth central cluster angles and subray angles, respectively. Similarly, $\chi_c^\textrm{el}$ and $\chi_s^\textrm{el}$ are the equivalents in elevation. In \cite{JSTSP} the convergence of these summations is shown to be extremely rapid, hence the bounds of the summations can be truncated to a reasonable number of terms without significant loss of accuracy.
\subsubsection{HURA}\label{ss:HURA_derivation}
%{\color{blue}\begin{itemize}
%	\item reference JSTSP for Kc
%	\item solve Ks
%\end{itemize}}
%
%Using the results from \cite{JSTSP}, we have:
%\begin{align}\label{eq:Kc_HURA}
%K_c^\textrm{HURA} &= 
%\sum_{m_x,m'_x}^{Mx-1}\sum_{m_y,m'_y}^{My-1}\sum_{n=-\infty}^\infty\sum_{n'=-\infty}^\infty\\\notag
%&(-1)^{p(n')}\chi_c^\textrm{az}(n')\chi_s^\textrm{az}(n')\chi_c^\textrm{el}(2n)\chi_s^\textrm{el}(2n)e^{jn'\alpha}I(n,n'),
%\end{align}
%where $z_1 = 2\pi d_x (m_x-m'_x)$, $z_2 = 2\pi d_y(m_y-m'_y)$, $z_T = \sqrt{z_1^2 + z_2^2}$ and $\alpha = \tan^{-1}(z_1/z_2)$ and we have the definitions
%\begin{align}\label{eq:I_c_HURA}
%I_c^\textrm{HURA}(n,n') = J_{\frac{|n|}{2}-n'}\left(\frac{z_T}{2}\right)J_{\frac{|n|}{2}+n'}\left(\frac{z_T}{2}\right)
%\end{align}
%and
%\begin{align}
%	\rho(x) = 
%	\begin{cases}
%		x & x<0\\
%		0 & x\geq0\\
%	\end{cases}.
%\end{align}
%We use the notation $\sum_{i,j}^{I-1}$ in place of $\sum_{i = 0}^{I-1}\sum_{j = 0}^{I-1}$.
%%
Substituting \eqref{eq:HURA_steering_vector} into \eqref{eq:Kc_definition} we have:
\begin{align}\label{eq:Kc_HURA_form}
&\begin{aligned}
K_c^\textrm{HURA} &= \frac{1}{M^2}\sum_{m_x,m'_x}^{M_x-1}\sum_{m_y,m'_y}^{M_y-1}\\
&\mathbb{E}\big[e^{j2\pi d_x (m_x-m'_x)(\sin\theta_{c,s}^{(l)}\cos\phi_{c,s}^{(l)}-\sin\theta_{c',s'}^{(l')}\cos\phi_{c',s'}^{(l')})}\\
&\times e^{j2\pi d_y(m_y-m'_y)(\sin\theta_{c,s}^{(l)}\sin\phi_{c,s}^{(l)}-\sin\theta_{c',s'}^{(l')}\sin\phi_{c',s'}^{(l')})}\big]
\end{aligned}\notag\\
&\begin{aligned}
=\frac{1}{M^2}\sum_{m_x,m'_x}^{M_x-1}\sum_{m_y,m'_y}^{M_y-1}|\mathbb{E}[e^{(j2\pi\sin\theta[z_1\cos\phi+z_2\sin\phi])}]|^2
\end{aligned}\notag\\
&\begin{aligned}
=\frac{1}{M^2}\sum_{m_x,m'_x}^{M_x-1}\sum_{m_y,m'_y}^{M_y-1}|\mathbb{E}[e^{(jz_T\sin\theta\sin(\phi + \alpha))}]|^2
\end{aligned}
\end{align}
with $z_1 = d_x(m_x-m'_x)$, $z_2 = d_y(m_y-m'_y)$, $z_T = 2\pi\sqrt{z_1^2+z_2^2}$, and $\alpha = \tan^{-1}(z_1/z_2)$.
From \cite{JSTSP} we have
\begin{align}\label{eq:Kc_HURA}
K_c^\textrm{HURA} &= 
\sum_{m_x,m'_x}^{M_x-1}\sum_{m_y,m'_y}^{M_y-1}\sum_{n=-\infty}^\infty\sum_{n'=-\infty}^\infty\\\notag
&(-1)^{\rho(n')}\chi_c^\textrm{az}(n')\chi_s^\textrm{az}(n')\chi_c^\textrm{el}(2n)\chi_s^\textrm{el}(2n)e^{jn'\alpha}\zeta_c(n,n'),
\end{align}
where $\zeta_c(n,n')=J_{|n|/2-n'}\left(z_T/2\right)J_{|n|/2+n'}\left(z_T/2\right)$ and $\rho(x) = \min(x,0)$. To solve for $K_s^\textrm{HURA}$ we substitute the HURA steering vector from \eqref{eq:HURA_steering_vector} into \eqref{eq:Ks_definition} and obtain 
\begin{align}
%&\begin{aligned}
&K_s^\textrm{HURA} = 
%\frac{1}{M^2}\sum_{m_x,m'_x}^{M_x-1}\sum_{m_y,m'_y}^{M_y-1}&\mathbb{E}[e^{j2\pi d_x (m_x-m'_x)\sin\theta_{c,s}^{(l)}\cos\phi_{c,s}^{(l)}}\\
%&\times e^{-j2\pi d_x (m_x-m'_x)\sin\theta_{c,s'}^{(l')}\cos\phi_{c,s'}^{(l')}}\\
%&\times e^{j2\pi d_y (m_y-m'_y)\sin\theta_{c,s}^{(l)}\sin\phi_{c,s}^{(l)}}\\
%&\times e^{-j2\pi d_y (m_y-m'_y)\sin\theta_{c,s'}^{(l')}\sin\phi_{c,s'}^{(l')}}]\\
%\end{aligned}\notag\\
\frac{1}{M^2}\sum_{m_x,m'_x}^{M_x-1}\sum_{m_y,m'_y}^{M_y-1}\\\notag
&\mathbb{E}\Big[e^{j2\pi\sin\theta_{c,s}^{(l)}(d_x (m_x-m'_x)\cos\phi_{c,s}^{(l)}+ d_y (m_y-m'_y)\sin\phi_{c,s}^{(l)})}\\\notag
&\times e^{-j2\pi \sin\theta_{c,s'}^{(l')}(d_x (m_x-m'_x)\cos\phi_{c,s'}^{(l')} + d_y (m_y-m'_y)\sin\phi_{c,s'}^{(l')})}\Big].
\end{align}
Using basic trigonometric results, this becomes
\begin{align*}
K_s^\textrm{HURA} =&\frac{1}{M^2}\sum_{m_x,m'_x}^{M_x-1}\sum_{m_y,m'_y}^{M_y-1}\numberthis\label{eq:Ks_HURA_form}\\
&\mathbb{E}\Big[e^{j z_T[\sin(\theta_{c,s}^{(l)})\sin(\phi_{c,s}^{(l)} + \alpha)-\sin(\theta_{c,s'}^{(l')})\sin(\phi_{c,s'}^{(l')} + \alpha)]}\Big].
\end{align*}
We provide an analytical expression for $K_s^\textrm{HURA}$ in Lemma \ref{res:HURA_Ks}.
\newtheorem{lemma}{Lemma}
\begin{lemma}\label{res:HURA_Ks}
	For a HURA, $K_s$ is given by
	\begin{align*}
	K_s^\textrm{HURA} &=\frac{1}{4M^2}\sum_{m_x,m'_x}^{M_x-1}\sum_{m_y,m'_y}^{M_y-1}\sum_{n=-\infty}^{\infty}\sum_{n'=-\infty}^{\infty}\sum_{k=-\infty}^{\infty}\sum_{k'=-\infty}^{\infty}\\\notag
	&\chi_c^\textrm{az}(n+n')\chi_s^\textrm{az}(n)\chi_s^\textrm{az}(n')\chi_c^\textrm{el}(k+k')\chi_s^\textrm{el}(k)\chi_s^\textrm{el}(k')\\\notag
	&\times e^{j(n+n')\alpha}e^{j\frac{\pi}{2}(k+k')}\zeta_s(n,n',k,k',z_T),\numberthis\label{eq:Ks_HURA}
	\end{align*}
	with
	\begin{align*}
	&\zeta_s(n,n',k,k',z_T)\\
	&= e^{-j(k+k')/(2\pi)}J_{\frac{|n|}{2}-\frac{k}{2}}\left(-z_T/2\right)J_{\frac{|n|}{2}+\frac{k}{2}}\left(-z_T/2\right)\\\notag
	&\times (-1)^{\rho(n)+\rho(n')}J_{\frac{|n'|}{2}-\frac{k'}{2}}\left(z_T/2\right)J_{\frac{|n'|}{2}+\frac{k'}{2}}\left(z_T/2\right).\numberthis
	\end{align*}
\end{lemma}

\begin{IEEEproof}
	The proof is given in  the Appendix.
%	~\ref{app:HURA_Ks}.
\end{IEEEproof}

\subsubsection{UCA}\label{ss:UCA_derivation}
%{\color{blue}\begin{itemize}
%	\item solve Kc
%	\item solve Ks
%\end{itemize}}
%
Substituting the UCA steering vector from \eqref{eq:UCA_steering_vector} into \eqref{eq:Kc_definition} and \eqref{eq:Ks_definition} we have
\begin{align*}
&\begin{aligned}
K_c^\textrm{UCA} =\frac{1}{M^2}\mathbb{E}\Bigg[|\sum_{m = 0}^{M-1}&e^{-j\frac{\pi d_r}{\sin(\pi/M)}\sin\theta_{c,s}^{(l)}\cos(\phi_{c,s}^{(l)}-\psi_m)}\\\notag
&\times e^{j\frac{\pi d_r}{\sin(\pi/M)}\sin\theta_{c',s'}^{(l')}\cos(\phi_{c',s'}^{(l')}-\psi_m)}|^2\Bigg]
\end{aligned}\\
&\begin{aligned}
=\frac{1}{M^2}\hspace{-0.1cm}\sum_{m,m'}^{M-1}&|\mathbb{E}[e^{\frac{j\pi d_r}{\sin(\pi/M)}\sin\theta(\cos(\phi-\psi_m)-\cos(\phi-\psi_{m'}))}]|^2
\end{aligned}\\
&\begin{aligned}
=\frac{1}{M^2}\sum_{m,m'}^{M-1}&|\mathbb{E}[e^{jz_T'\sin\theta\sin(\phi + \alpha')}]|^2,
\end{aligned}\numberthis\label{eq:Kc_UCA_form}
\end{align*}
and
\begin{align}\label{eq:Ks_UCA_form}
&\begin{aligned}
K_s^\textrm{UCA}\hspace{-0.1cm}= &\frac{1}{M^2}\hspace{-0.15cm}\sum_{m,m'}^{M-1}\hspace{-0.1cm}\mathbb{E}[e^{\frac{-j\pi d_r}{\sin(\pi/M)}\sin\theta_{c,s}^{(l)}[\cos(\phi_{c,s}^{(l)}-\psi_m)-\cos(\phi_{c,s}^{(l)}-\psi_{m'})]}\\\notag
&\times e^{j\frac{\pi d_r}{\sin(\pi/M)}\sin\theta_{c,s'}^{(l')}[\cos(\phi_{c,s'}^{(l')}-\psi_m)-\cos(\phi_{c,s'}^{(l')}-\psi_{m'})]}]
\end{aligned}\\
&\begin{aligned}
=\frac{1}{M^2}\hspace{-0.2cm}\sum_{m,m'}^{M-1}&\mathbb{E}[e^{jz_T'\sin\theta_{c,s}^{(l)}\sin(\phi_{c,s}^{(l)} + \alpha')}e^{-jz_T'\sin\theta_{c,s'}^{(l')}\sin(\phi_{c,s'}^{(l')} + \alpha')}],
\end{aligned}
\end{align}
with $z_1' = \cos\psi_{m'} - \cos\psi_m$, $z_2' = \sin\psi_{m'}-\sin\psi_m$, $z_T' = \frac{\pi d_r}{\sin(\pi/M)}\sqrt{z_1'^2 + z_2'^2}$, and $\alpha' = \tan^{-1}(z_1'/z_2')$.
Here, we see that $K_c^\textrm{UCA}$ in \eqref{eq:Kc_UCA_form} and $K_s^\textrm{UCA}$ in \eqref{eq:Ks_UCA_form} are identical in form to $K_c^\textrm{HURA}$ and $K_s^\textrm{HURA}$ in \eqref{eq:Kc_HURA_form} and \eqref{eq:Ks_HURA_form}. Hence we provide analytical expressions for $K_c^\textrm{UCA}$ and $K_s^\textrm{UCA}$ in Lemma \ref{lemma:UCA_results}.
\begin{lemma}\label{lemma:UCA_results}
	$K_c^\textrm{UCA}$ and $K_s^\textrm{UCA}$ are given by \eqref{eq:Kc_HURA} and \eqref{eq:Ks_HURA} with $z_1=z_1', z_2=z_2', z_T=z_T', \alpha=\alpha'$.
\end{lemma}

\section{Large System Convergence Rate}\label{s:kappa_decay}
We now examine the rate of decay of $\kappa^\textrm{FP}$ for a ULA in the large system limit using the methodology from \cite{jasmine_icc19}, extended to accommodate user cluster sharing. From \eqref{eq:kappa_FP_3}, this decay rate is determined by the decay rates of $K_c$ and $K_s$. For $K_c^\textrm{ULA}$ we begin by reducing \eqref{eq:Kc_ULA} to a single summation of the form
\begin{align*}
K_c^\textrm{ULA} &= \frac{1}{M}\left(1 + 2\sum_{m=1}^{M-1}\left(1 - \frac{m}{M}\right)\left|\mathbb{E}\left[e^{-j2\pi d_x m\sin\phi}\right]\right|^2\right)\\\notag
&= \frac{1}{M}\left(1 + 2\nu\right).\numberthis\label{eq:K_c_ULA_breakdown}
\end{align*}
In \cite{2019_TWC_SLi}, it is shown that $\nu$ is $\mathcal{O}(\log M)$ in most situations, but in the absence of end-fire radiation $\nu$ is $\mathcal{O}(1)$.
Hence, $K_c^\textrm{ULA}$ decays as $\mathcal{O}\left(1/M\right)\leq\mathcal{O}\left(K_c^\textrm{ULA}\right)\leq\mathcal{O}\left(\log M/M\right)$.
For $K_s^\textrm{ULA}$, we have the form
\begin{align*}\label{eq:K_s_ULA_breakdown}
K_s^\textrm{ULA} =\frac{1}{M^2}\sum_{m,m'}^{M - 1}\mathbb{E}\bigg[&e^{-j2\pi d_x\left(m-m'\right)\left(\sin\left(\phi_{c,s}^{(l)}\right) - \sin\left(\phi_{c,s'}^{(l')}\right)\right)}\bigg]\\\notag
\begin{split}=\mathbb{E}_{\phi}\bigg[\frac{1}{M^2}\sum_{m,m'}^{M - 1}\bigg|\mathbb{E}_\Delta\bigg[&e^{-j2\pi d_x\left(m-m'\right)\left(\sin\left(\phi_c + \Delta_{c,s}^{(l)}\right)\right)}\bigg]\bigg|^2\bigg]\end{split}.\numberthis
\end{align*}
The contents of the expectation $\mathbb{E}_{\phi}[\cdot]$ in \eqref{eq:K_s_ULA_breakdown} are identical in form to \eqref{eq:Kc_ULA} with an additional angular offset $\phi_c$ which is constant relative to the expectation. By the same process used to analyse the decay rate of $K_c^\textrm{ULA}$, we find that $K_s^\textrm{ULA}$ also decays as $\mathcal{O}(1/M)\leq\mathcal{O}(K_s^\textrm{ULA})\leq\mathcal{O}(\log M/M)$. This implies the decay rate of $\kappa^\textrm{FP}$ is $\mathcal{O}(1/M)\leq\mathcal{O}(\kappa^\textrm{FP})\leq\mathcal{O}(\log M/M)$.
%\subsection{FP}
%\subsubsection{ULA}
%\subsubsection{HURA}
%\subsubsection{UCA}
%\subsection{CH}
%\subsubsection{ULA}
%\subsubsection{HURA}
%\subsubsection{UCA}
\section{Numerical Results}\label{results}
%
%% trim and size all figures equally:
\newcommand{\myincludegraphics}{\includegraphics[trim=1cm 0.04cm 1.6cm 0.66cm, clip=true, width=1\columnwidth]}
\newcommand{\raisecapt}{\vspace{-0.8cm}}
\setlength{\textfloatsep}{5.0pt plus 1.0pt minus 4.0pt}
\setlength{\textfloatsep}{10.0pt plus 1.0pt minus 2.0pt}
\setlength{\textfloatsep}{10.0pt plus 1.0pt minus 2.0pt}
%%\textfloatsep — distance between floats on the top or the bottom and the text;
%%\floatsep — distance between two floats;
%%\intextsep — distance between floats inserted inside the page text (using h) and the text proper.

%{\color{blue}\begin{itemize}
%	\item explain low angular spread and high angular spread scenarios and where they come from
%\end{itemize}}
This section presents numerical results and a discussion of the trends observed in $\kappa^\textrm{FP}$ for different antenna topologies and varying levels of cluster sharing. Recall that higher levels of $\kappa^\textrm{FP}$ indicate a slower convergence to FP.

%Hence, lower values of $\kappa^\textrm{FP}$ and $\kappa^\textrm{CH}$ are generally speaking indicative of higher performance.\newline
Table \ref{parameters} gives the parameters considered. The angular spread values for Scenarios 1 and 2 were obtained from \cite{sangodoyin_cluster_2018} and \cite{3GPP}, respectively, and a square HURA is considered. Central cluster angles are Gaussian distributed in azimuth and Laplacian distributed in elevation while subray angles are Laplacian distributed in both cases, in accordance with \cite{sangodoyin_cluster_2018} and \cite{3GPP}.
\begin{table}[ht]
	\caption{Parameters for Numerical Results}
	\centering
	\label{parameters}
	\begin{tabular}{c|c}
		\hline\hline
		\textbf{Parameter} & \textbf{Values}\\
		\hline
		antenna spacing, $d_r$, $d_x$ and $d_y$ & 0.5\\
		\hline
		Azimuth values&\\
		cluster angle mean, $\mu_{\mathrm{c}}$ & $0\degree$\\
		cluster angle variance, $\sigma_{\mathrm{c}}^2$, (Scen. 1, Scen. 2)& {$(14.4\degree)^2$, $(31.64\degree)^2$}\\
		subray angle variance, $\sigma^2_{\mathrm{s}}$, (Scen. 1, Scen. 2) & {$(6.24\degree)^2$, $(24.25\degree)^2$}\\ [1ex]
		\hline
		Elevation values&\\
		cluster angle mean, $\hat{\mu}_{\text{c}}$ & $90\degree$\\
		cluster angle variance, $\hat{\sigma}_{\mathrm{c}}^2$, (Scen. 1, Scen. 2)& {$(1.9\degree)^2$, $(6.12\degree)^2$}\\
		subray angle variance, $\hat{\sigma}^2_{\mathrm{s}}$, (Scen. 1, Scen. 2) & {$(1.37\degree)^2$, $(1.84\degree)^2$}\\ [1ex]
		\hline
	\end{tabular}
\end{table}

%\begin{table}[ht]
%	\caption{Cluster Sharing Probabilities}
%	\vspace{-0.3cm}
%	\centering
%	\begin{tabular}{|c|c|c|}
%		\hline
%		\textbf{$C_t$}& \textbf{$p_\textrm{sh}$}& \textbf{$r_\textrm{VR}:r_\textrm{cell}$}\\
%		\hline\hline
%		1&1&??\\
%		3&0.33&??\\
%		10&0.1&??\\
%		20&0.05&??\\
%		100&0&??\\
%		\hline
%	\end{tabular}
%	\label{probabilities}
%\end{table}

%%%%%%%%%%%% ZF SNR CDF %%%%%%%%%%%%%%%

\begin{figure}[ht]
	\centering
	\myincludegraphics
	{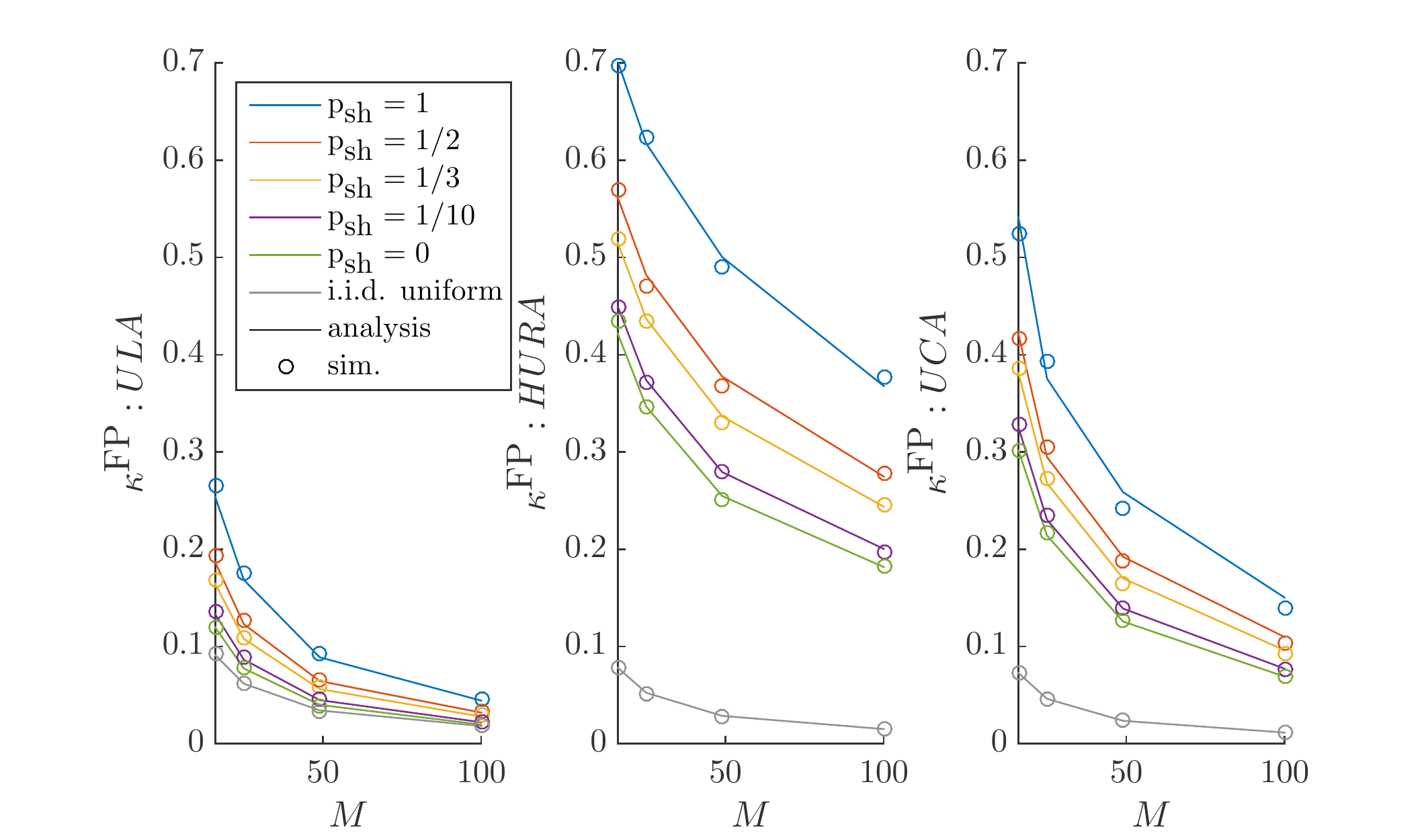}
	\raisecapt\caption{$\kappa^\textrm{FP}$ vs $M$ for all antenna configurations: low spread (Scen. 1).}
	\label{fig:kappa_FP_lowspread}
\end{figure}

\begin{figure}[ht]
	\centering
	\myincludegraphics
	{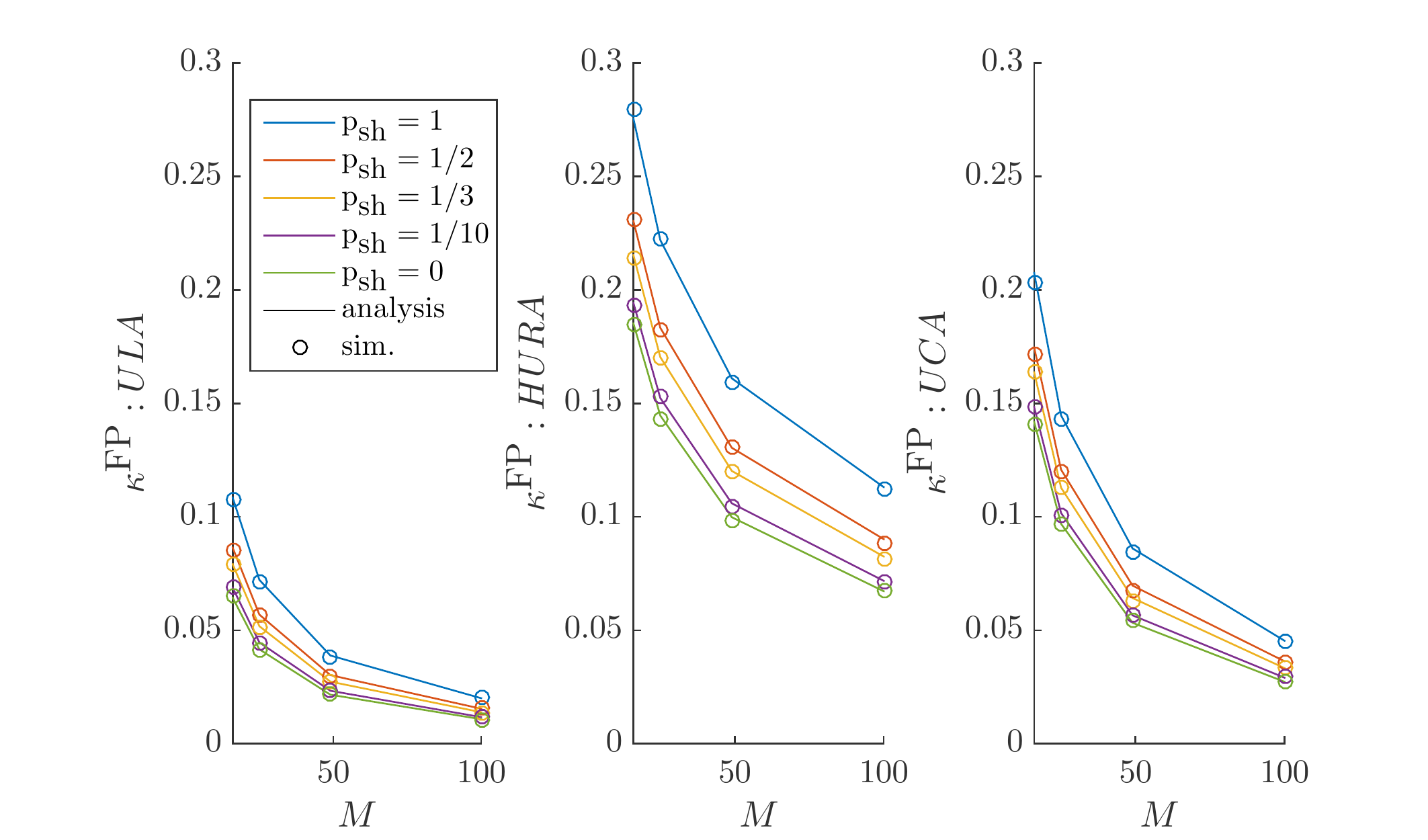}
	\raisecapt\caption{$\kappa^\textrm{FP}$ vs $M$ for all antenna configurations: high spread (Scen. 2).}
	\label{fig:kappa_FP_highspread}
\end{figure}

Figs.~\ref{fig:kappa_FP_lowspread} and ~\ref{fig:kappa_FP_highspread} examine the MSD from FP  for Scenarios 1 and 2, respectively. For comparison, we also include results for i.i.d. spherically uniform ray angles in Fig.~\ref{fig:kappa_FP_lowspread}. We evaluate $\kappa^\textrm{FP}$ by simulation using \eqref{eq:kappa_FP_2}{\footnote{As discussed in Sec.~\ref{ss:kappa_FP}, $\kappa^\textrm{FP}$ is independent of $C$ and $S$. However, for the purpose of simulating channels, we use $C=1$ and $S=16$.}} with $3\times10^4$ samples, and by analysis using the results in Sec.~\ref{s:derivations} with the characteristic functions of the Gaussian and Laplacian angular variables given by $\chi_c^\textrm{az}(n) = \exp(jn\mu_c - n^2\sigma_c^2/2)$, $\chi_s^\textrm{az}(n) = (1 + n^2\sigma_s^2)^{-1}$, $\chi_c^\textrm{el}(n) = (1 + n^2\hat{\sigma}_c^2)^{-1}$, and $\chi_s^\textrm{el}(n) = (1 + n^2\hat{\sigma}_s^2)^{-1}$.
The first trend observed from Figs. ~\ref{fig:kappa_FP_lowspread} and ~\ref{fig:kappa_FP_highspread} is the significant increase of $\kappa^\textrm{FP}$ with increased cluster sharing. When the channels from two users share common clusters, some of the sub-rays from the different users will be centred around the same central angle. Hence, the channels become more correlated which increases $\kappa^\textrm{FP}$. Values of $p_\textrm{sh}$ as small as $1/3$ can increase $\kappa^\textrm{FP}$ by up to 25\% relative to $p_\textrm{sh}=0$ - a significant effect considering that the ``significance of common clusters" has been observed to be as high as 95\% in indoor measurements \cite{poutanen2012multi}.

Comparing the size of $\kappa^\textrm{FP}$ in Figs.~\ref{fig:kappa_FP_lowspread} and ~\ref{fig:kappa_FP_highspread}, we observe that a larger angular spread decreases $\kappa^\textrm{FP}$. Increasing the angular spread and thus the channel diversity decreases the average similarity between the channels of two different users. This in turn reduces the inner product in \eqref{eq:kappa_FP_2} and, consequently, $\kappa^\textrm{FP}$.

In Figs.~\ref{fig:kappa_FP_lowspread} and ~\ref{fig:kappa_FP_highspread}, we see that the ULA provides the smallest distance to FP, followed by the UCA, and finally the HURA. We attribute this to the corresponding azimuth footprint reduction of these topologies. As seen in the angular parameters from measurements (see Table~\ref{parameters}), the majority of the angular diversity is contained in the azimuth plane. Increasing the azimuth footprint of the antenna topology therefore provides more spatial diversity, hence decreasing $\kappa^\textrm{FP}$.

%\begin{itemize}
%	\item How should we talk about FP? Orthogonality of user channels? Passive elimination of inter-user interference?
%	\item greater "distance from FP" for lower angle spread
%	\item\begin{itemize}
%		\item lower angle spread = rays closer together/more similar on average
%		\item closer together/more similar = larger average inner product = higher kappa
%	\end{itemize}
%	\item greater difference from FP for larger percentage of cluster sharing
%	\item\begin{itemize}
%		\item shared cluster causes sub-rays from two different users to be centred around the same set of central angles
%		\item forces rays together - channels are now highly correlated
%		\item greater inner product = higher kappa
%	\end{itemize}
%	\item distance from FP: HURA>UCA>ULA
%	\item\begin{itemize}
%		\item ULA converges quickest because largest azimuth footprint
%		\item HURA has a larger azimuth footprint than UCA, however the spread of antennas is more homogeneous when seen from different perspectives in the azimuth plane. There exist several angles from which more than half of the available antennas are aligned directly behind other antennas at uniform intervals, diminishing diversity. UCA has, at worst, half the antennas aligned behind other antennas, and at non-uniform spacing. There will always be at least $M/2$ apparent antennas along the azimuth footprint. 
%	\end{itemize}
%\end{itemize}
The insight regarding the azimuth footprint of the topologies agrees with previous results for ray-based channels found using simulation in \cite{aslam_performance_2019}, but contradicts previous analytical results in \cite{wu_favorable_2017}. The latter finds that the antenna topology has little effect on the distance from FP. We conclude that the use of uniform angular distributions in \cite{wu_favorable_2017} obscures the effects of the antenna topology with realistic angular distributions. This is illustrated by the i.i.d. spherically uniform results in Fig.~\ref{fig:kappa_FP_lowspread} which show little variation across topologies. In addition, with spherically uniform rays the MSD from FP is highest for the ULA and lowest for the UCA (see Fig.~\ref{fig:kappa_FP_lowspread} and \cite{wu_favorable_2017}), whereas for Scenarios 1 ad 2, the MSD is highest for the HURA and lowest for the ULA. This change in ordering is best explained in terms of a ULA. With Scenarios 1 and 2, the ray azimuth distributions are concentrated near broadside. From this angle, the topology has a wide azimuth footprint. However, for the spherical uniform case, there is a wide range of ray angles from which the footprint of the topology appears much narrower. Hence, we see that the use of unrealistic angular distributions drastically alters both the size and nature of the topology effects on FP behavior.
%\begin{figure}[ht]
%	\centering
%	\myincludegraphics
%	%\includegraphics[trim=1cm 0.04cm 1.6cm 0.78cm, clip=true, width=1\columnwidth] 
%	{sig_vs_M_sc.pdf}
%	\raisecapt\caption{$\kappa^\textrm{CH}$ vs $M$ for all configurations; both scenarios.}
%	\label{fig:kappa_CH}
%\end{figure}

The following insights can be gained from these observations. For moderate values of $M$, for which the spatial resolution is sufficient to distinguish clusters, but not subrays, FP behavior is determined by diversity amongst clusters. Thus, user cluster sharing diminishes performance for moderate $M$, as there are fewer degrees of freedom to separate UEs. As $M$ increases further, the spatial resolution becomes sufficiently fine to distinguish subrays. In this "subray-dominated region", user cluster sharing has less of an effect because all subrays deliver nearly orthogonal steering vectors. Thus, we see the values of $\kappa^\textrm{FP}$ for different sharing probabilities converge in Figs.~\ref{fig:kappa_FP_lowspread} and ~\ref{fig:kappa_FP_highspread}: in the limit $M\to\infty$, cluster sharing does not play a role. The transition region from cluster- to subray-dominated resolution is not only determined by $M$, but also the array shape. The azimuthal resolution of a HURA is much less than a UCA, and further, a ULA for same element spacing. Hence the antenna topologies with larger azimuth footprints will enter the subray-dominated region at lower values of $M$.

The observed trends intuitively align with the common understanding of massive MIMO systems' reliance on diverse channel conditions. Wider angular spreads increase diversity, and a wider azimuth footprint provides more spatial information to capture diversity. This facilitates the natural ability of MIMO to passively orthogonalize user channels. Cluster sharing diminishes diversity, impairing this functionality.

\begin{figure}[ht]
	\centering
	\myincludegraphics
	{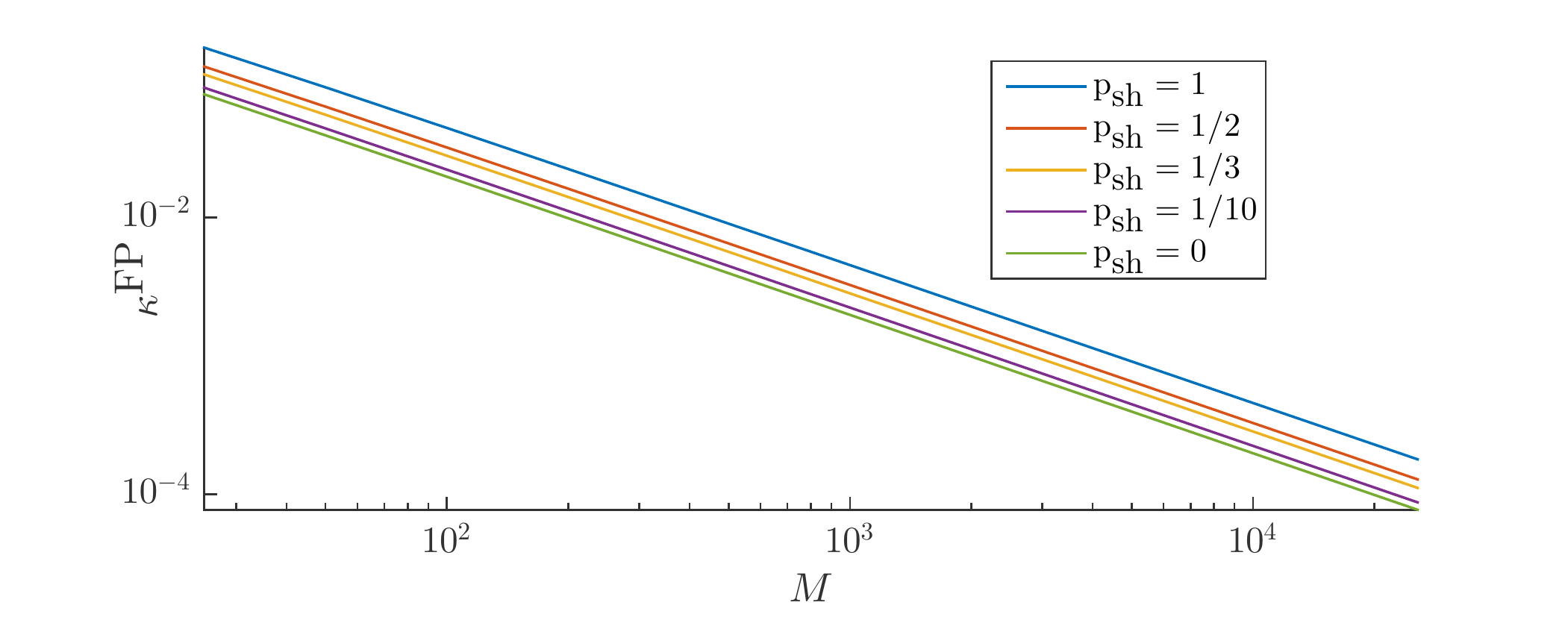}
	\raisecapt\caption{Decay rate of $\kappa^\textrm{FP}$ for ULA in the large system limit (Scen. 1).}
	\label{fig:kappa_FP_largeM}
\end{figure}
Finally, in Fig.~\ref{fig:kappa_FP_largeM} we examine the large-system decay rate of $\kappa^\textrm{FP}$ for a ULA in Scenario 1. Here, we plot $\kappa^\textrm{FP}$ with both $x-$ and $y-$ axes on a logarithmic scale. From Sec.~\ref{s:kappa_decay} we know that $\mathcal{O}(1/M)\leq\mathcal{O}(\kappa^\textrm{FP})\leq\mathcal{O}(\log(M)/M)$.
%On a log scale, this becomes $\mathcal{O}(-\log M)\leq\mathcal{O}(\log\kappa^\textrm{FP})\leq\mathcal{O}(\log(\log M)-\log M)$. 
For the system simulated in Fig.~\ref{fig:kappa_FP_largeM} the value of $\mathcal{O}(\kappa^\textrm{FP})$ is $\mathcal{O}(\log(M)/M)$. Hence a plot of $\log(\kappa^\textrm{FP})$ vs $\log(M)$ follows an approximately linear trend for large $M$ as $\log(\log(M))$ is extremely slowly varying. This pattern is verified in Fig.~\ref{fig:kappa_FP_largeM}.
%Hence a plot of $\log(\kappa^\textrm{FP})$ vs $\log(M)$ follows a roughly linear trend for large $M$, as can be seen in Fig.~\ref{fig:kappa_FP_largeM}. 
%%%%%%%%%%%% CONCLUSION %%%%%%%%%%%%%%%
%
\section{Conclusions}\label{conclusion}
We prove the existence of FP for a ULA and HURA topology for a ray-based channel model with user cluster sharing, and provide conjecture for FP existence for a UCA. We provide analytical expressions for the distance from FP for a finite-antenna system with each of these topologies, and prove that the distance from FP decays at a rate between $\mathcal{O}(1/M)$ and $\mathcal{O}(\log M/M)$ in the large system limit with cluster sharing, as is the case without cluster sharing. We use our results to identify the detrimental effect of user cluster sharing on FP behavior, and determine that FP behavior is best facilitated by topologies with larger azimuth footprints.
%
%
%
%%%%%%%%%%%% APPENDIX %%%%%%%%%%%%%%%
%
\appendix[Proof of Lemma \ref{res:HURA_Ks}]\label{app:HURA_Ks}
Writing $\phi_{c,s}^{(l)} = \phi_c + \Delta_{c,s}^{(l)}$ and $\theta_{c,s}^{(l)} = \theta_c + \delta_{c,s}^{(l)}$, \eqref{eq:Ks_HURA_form} becomes
\begin{align*}
K_s^\textrm{HURA} =\frac{1}{M^2}&\sum_{m_x,m'_x}^{M_x-1}\sum_{m_y,m'_y}^{M_y-1}\mathbb{E}[e^{jz_T\sin(\theta_c + \delta_{c,s}^{(l)})\sin(\phi_c + \Delta_{c,s}^{(l)} + \alpha)}\\
&\times e^{-jz_T\sin(\theta_c + \delta_{c,s'}^{(l')})\sin(\phi_c + \Delta_{c,s'}^{(l')} + \alpha)}].\numberthis\label{app:Ks_HURA_form}
\end{align*}
We first take the average over azimuth subray offsets using the result for $\mathbb{E}_{\Delta}[e^{jB\sin(\Delta + C)}]$ from App. D of \cite{JSTSP} and obtain
\begin{align*}
K_s^\textrm{HURA} =&\frac{1}{M^2}\sum_{m_x,m'_x}^{M_x-1}\sum_{m_y,m'_y}^{M_y-1}\\
&\mathbb{E}_{\phi, \theta, \delta}\Big[\mathbb{E}_{\Delta}[e^{j2\pi z_T\sin(\theta_c + \delta_{c,s}^{(l)})\sin(\phi_c + \Delta_{c,s}^{(l)} + \alpha)}\\
&\times e^{-j2\pi z_T\sin(\theta_c + \delta_{c,s'}^{(l')})\sin(\phi_c + \Delta_{c,s'}^{(l')} + \alpha)}]\Big]\\
=&\frac{1}{M^2}\sum_{m_x,m'_x}^{M_x-1}\sum_{m_y,m'_y}^{M_y-1}\numberthis\\
&\mathbb{E}_{\phi, \theta, \delta}\Big[\hspace{-0.2cm}\sum_{n=-\infty}^{\infty}\hspace{-0.2cm}e^{jn(\phi_c + \alpha)}\chi_s^\textrm{az}(n)J_n(z_T\sin(\theta_c+\delta_{c,s}^{(l)}))\\
&\times\hspace{-0.3cm}\sum_{n'=-\infty}^{\infty}\hspace{-0.2cm}e^{jn'(\phi_c + \alpha)}\chi_s^\textrm{az}(n')J_{n'}(-z_T\sin(\theta_c+\delta_{c,s'}^{(l')}))\Big].
\end{align*}
Evaluating the expectation over azimuth central cluster angles using the definition of the characteristic function $\chi_c^\textrm{az}$, we have
\begin{align*}
%\begin{aligned}
K_s^\textrm{HURA} &=\frac{1}{M^2}\sum_{m_x,m'_x}^{M_x-1}\sum_{m_y,m'_y}^{M_y-1}\numberthis\\
&\sum_{n=-\infty}^{\infty}\sum_{n'=-\infty}^{\infty}e^{j(n+n')\alpha}\chi_c^\textrm{az}(n+n')\chi_s^\textrm{az}(n)\chi_s^\textrm{az}(n')\\
&\times\mathbb{E}_{\theta, \delta}[J_n(z_T\sin(\theta_c\hspace{-0.05cm}+\hspace{-0.05cm}\delta_{c,s}^{(l)}))J_{n'}(-z_T\sin(\theta_c\hspace{-0.05cm}+\hspace{-0.05cm}\delta_{c,s'}^{(l')}))].\\
%\end{aligned}
\end{align*}
Finally, the expectation over elevation subray offsets is computed using the same technique used in \cite{chelsea_icc19} and \cite{JSTSP} where the angular PDF is replaced with its Fourier series. This gives:
\begin{align}
\mathbb{E}_{\theta, \delta}&[J_n(z_T\sin(\theta_c+\delta_{c,s}^{(l)}))J_{n'}(-z_T\sin(\theta_c+\delta_{c,s'}^{(l')}))]\\\notag
&=\frac{1}{4\pi^2}\sum_{k=-\infty}^{\infty}\sum_{k'=-\infty}^{k=\infty}\chi_s^\textrm{el}(k)\chi_s^\textrm{el}(k')\mathbb{E}_{\theta}[e^{j(k+k')\theta}]\\\notag
&\times\int_{0}^{\pi}e^{-jkx}J_n(z_T\sin x)dx\int_{0}^{\pi}e^{-jk'x}J_{n'}(-z_T\sin x)dx.
\end{align}
We evaluate the angular average in elevation over the range $[0,\pi]$. This approximation covers the bulk of the PDF which is centred on $\pi/2$ and typically has a standard deviation less than $10^o$ from measurement \cite{3GPP,sangodoyin_cluster_2018}.  Using \cite[Equations 6.681.8 and 6.681.9]{gradshteyn2007} we have the desired solution.

\bibliographystyle{IEEEtran}
%\bibliography{IEEEabrv,bibliography}
\bibliography{bibliography}
\end{document}